# How does heat propagate in liquids?

## *Fabio Peluso*

*Leonardo SpA – Electronics Division – Defense Systems LoB*
*Via Monterusciello 75, 80078 Pozzuoli (NA) - Italy*
*mailto: fpeluso65@gmail.com.*

## Abstract

We proceed further in this paper to illustrate the consequences and implications of the Dual Model of Liquids (DML) by applying it to the heat propagation. Within the frame of the DML, propagation of thermal (elastic) energy in liquids is due to wave-packets propagation and their interaction with the material particles of the liquid, meant in the DML as aggregates of molecules swimming in an ocean of amorphous liquid. The liquid particles interact with the lattice particles, a population of elastic wave-packets, by means of an inertial force, exchanging with them energy and momentum. The hit particle relaxes at the end of the interaction giving the energy and momentum back to the system a step forward and a time lapse later, alike in a tunnel effect.

The tunnel effect and the duality of liquids are the new elements that suggest on a physical basis, for the first time, using a hyperbolic equation to describe the propagation of energy associated to the dynamics of wave-packets interaction with liquid particles. Although quantitatively relevant only in the transient phase, the additional term characterizing the hyperbolic equation, usually named the "memory term", is physically present also once the stationary is attained; it is responsible for dissipation in liquids and provides a finite propagation velocity for wave packets avalanches responsible in the DML for the heat conduction. The consequences of this physical interpretation of the "memory" term added to the Fourier law for the phononic contribution are discussed and compiled with numerical forecasts for the value of the memory term and with the conclusions of other works on the same topic.

## 1.    Introduction.

The paradox of the equations that describe the diffusion of energy, mass, or something else, in physics has been known for some time now. This paradox consists in the fact that the equations which mathematically describe the diffusion, being of the parabolic type (e.g. Fourier, Fick, Dufour), are characterized by an infinite diffusion velocity: if, at the initial instant, there is a disturbance on one edge of the system, this disturbance is instantaneously present "at the same time" in every other point of the system, although with different intensity. This paradox, it is equally known, is due to the fact that the parabolic equations normally used to describe the phenomenon of propagation of signals in material media are rather suitable to describe the phenomenon of diffusion, and as such are strictly valid only in stationary conditions, i.e. when every transient effect has been overcome in the dynamics of the system at microscopic level. It is also equally well known that the hyperbolic type equations are among those suitable for describing the propagation of mass and/or energy in a material medium even in the transient phase. In fact, these equations, also known as the "telegrapher' equation", when the proper initial and boundary conditions are given, provide solutions that have an initial undulatory nature without back front, present only in the transient, and become then of diffusive type, or parabolic, when the observation time of the phenomenon is much longer than the transient characteristic time, making the memory term negligible. Hyperbolic equations indeed allow to correctly describe those systems characterized by a "memory effect", i.e. traditionally those for which the response to a thermodynamic force comes later that the application of the force.

The physical reason for the above resides in the fact that the equilibrium state is not achieved instantaneously because thermal energy is carried microscopically by molecular interactions and proceeds with a finite velocity. With $J_q$ as the heat flux, $K$ the thermal conductivity and $T$ the temperature, there is a relaxation time $\tau$ during which Fourier's law

1. $\quad J_q = -K\nabla T$

does not hold. Equation (1) contains also another information, i.e. that if a uniform temperature gradient is established throughout a homogeneous fluid, then the heat flow is everywhere and every time proportional to the temperature gradient. Equations like (1) are very familiar in Irreversible Thermodynamics because they express the Onsager's theory property that the response of a system to an applied force is *simultaneous* with the application of the force. As a general rule, such *simultaneity* in a macroscopic theory turns out to be an approximation to a *causal* behavior, where the response to a force comes *after* the application of the force. To account for the *causality* behavior, it is necessary to modify the equation for heat conduction. Cattaneo [1] was the first to propose that a term linear in the derivative of the heat flux be added to Fourier law to account for the relaxation behavior. Cattaneo in his formulation adopted a physical argument due to Maxwell [2] and reused by Boltzmann [3], based on the collisional mechanisms working at molecular level. This formulation consisted in getting a modified form for the expression for the heat conductivity which, in turn, resulted in adding the term linear in the derivative of the heat flux in the Fourier law. Few years after Cattaneo, Vernotte [4-5], who seems not to know the Cattaneo's work [6], proposed the same equation to remove the paradox, although the term was added by him *ad hoc* into the Fourier equation on a pure mathematical basis. Whatever of the two formalisms one takes as example, that of Cattaneo or that of Vernotte, what matters is that Nettleton [7] has demonstrated that i) the so-called "telegrapher" equation may be viewed formally, from the point of view of the Onsager Irreversible Thermodynamics [8-9], as a force-flux equation linking two irreversible processes, and ii) a linear relaxation equation for liquids is consistent with the assumption that thermal energy is carried by elastic waves of very high frequencies. Very interestingly, Nettleton uses the Debye/Brillouin model for heat conduction [10-13], although he voluntary neglects the shear waves due to the supposed smallness of their contribution to the total balance.

The telegrapher equation was obtained [1-6] or used [14-19] by many Authors in the past to describe the heat conduction in liquids. In the same way, many Authors [16-19] have studied the diffusion of "something" in condensed media, but they were usually interested in the behavior in stationary conditions, so the paradox of the infinite velocity of energy diffusion has never emerged and discussed at physical level before Cattaneo. An historical excursus of the development of the Cattaneo equation for heat propagation is provided in the ***Appendix I***.

The mechanism classically believed at the base of heat conduction in fluids is that of molecular collisions: the thermal energy content is distributed among molecules proportionally to their degrees of freedom (DoF), the higher the energy content, the higher the number of DoF excited. By applying a heat flux to one edge of a fluid medium, the energy transferred to the system is distributed to the molecules from the hot to the cold side by means of avalanches of energetic particles that redistribute the excess of thermal energy by means of collisions. However such a mechanism has a weak point just in the quasi-instantaneous character of the collision among molecules, making the identification and the calculation of the relaxation time a hard exercise. Besides, it does not provide the physical mechanism requested for the telegrapher equation to work.

In a series of recent papers [20-21] we have presented a model of organization of the liquid state at mesoscopic level, dubbed Dual Model of Liquids (DML); it was even applied to calculate several liquid specific quantities, in particular the thermal conductivity, the order-of-magnitude of the relaxation times governing the interaction process and the liquid specific heat, finding good correspondence with the experimental data. In few words, DML assumes that liquid molecules are arranged at mesoscopic level on solid-like local lattices. Propagation of perturbations occurs at characteristic timescales typical of solids within these local domains of coherence, while mutual interactions of local clusters with inelastic wave-packets allow exchanging with them energy and momentum. The liquid is then a Dual System, the two subsystems being the *liquid particles* (i.e. the clusters of molecules) and the *lattice particles* (i.e. the wave-packets[1]). The purpose of this work is to show that if the propagation of heat in liquids is due to the mutual interactions of these two subsystems constituting the liquid, it is suitably described by a Cattaneo-like equation. To the best of the author's knowledge, this is the first time the telegrapher equation is proposed to be applied to the propagation of phonons in liquids as carriers of the thermal (and elastic) signal. Indeed it is just the presence of two interacting populations to justify the presence of the relaxation term in the equation for heat transport, that we propose to be linear in the relaxation time as early proposed by Cattaneo and Vernotte. The tunnel effect characterizing the interaction and the interaction itself are the physical mechanisms allowing for the interpretation of the heat transport to be of a dynamical nature, thus excluding also the dilemma of infinite diffusion velocity. The additional time derivative term characterizing the hyperbolic equations originates in the energy balance equation where the intrinsically negative term $\tau \dfrac{\partial J_q}{\partial t}$ is added to the Fourier law:

2.    $J_q + \tau \dfrac{\partial J_q}{\partial t} = -K\nabla T$ ;

it accounts for that part of the energy temporary stored into the system for the time interval $\tau$, i.e. the *relaxation time,* in a non propagating form. During such time-lapse, the linear relationship between the heat flux and the temperature gradient does not hold (even if the presence of $\tau \dfrac{\partial J_q}{\partial t}$ makes $J_q$ and $\nabla T$ no more proportional everywhere and every time).

---

[1] We will use through the manuscript the terms lattice-particle, wave-packet, phonon, collective excitation on one side, or liquid particle, molecular cluster and iceberg on the other, interchangeably, as synonyms.

The problem is certainly not new, many texts and scientific works have been written about (see Appendix I). However, they have mainly been focused on the mathematical problem, coming only to estimate on the one hand the extent of the "delay" in the propagation of the signal, i.e. the relaxation time, and on the other to assume that there should be a mechanism of "inelastic" type at its base. Nevertheless, the most important ingredient to complete the recipe has always been lacking, namely the identification of such a physical mechanism underlying the propagation phenomenon and therefore capable of allowing a calculation of relaxation times on a physical basis at the mesoscopic level. It is precisely in this framework that this work is placed, by providing the missing link through the DML.

Another important aspect correlated with the use of hyperbolic equations and typical of systems made of two interacting subsystems, is the emergence of a gap in the momentum space, usually known as the "*k-gap*" or *Gapped Momentum States* (GMS) [22-25]. We will show that the DML is even characterized by a *k-gap*, whose importance is crucial in the identification of the relaxation times typical of a dual system [20].

Here to follow, although what stated is equally valid both for mass or energy propagation, reference will preferentially be given to energy propagation in the form of heat. Any effect due to internal friction among molecules, i.e. viscosity, is neglected because the system is supposed at rest and in a stable configuration to avoid convection; in the same way, any effect of thermal expansion is even not considered. Finally, dispersionless systems are considered for simplicity, and any variation with temperature or pressure of the several quantities are neglected.

The paper is organized as follows. The paragraph 2.1 is dedicated to briefly recall the DML (the complete description is available elsewhere [20-21]). The paragraph 2.2 is dedicated to the derivation of the heat propagation equation in the form of the Cattaneo equation. In particular, the physical meaning of the relaxation time characteristic of the *lattice particle* ↔ *liquid particle* elementary interactions of DML is discussed. In the *Discussion* the consequences of the approach are illustrated as well as the physical interpretation of the several parameters characterizing the heat propagation equation. Finally, in the *Conclusions* future developments of the DML are anticipated. The Appendix I contains some interesting historical implications on the development and use of hyperbolic equations (or Cattaneo equations) to describe the heat propagation in condensed phases.

## 2. The Dual Model of Liquids and the heat propagation equation with "memory".

The Fourier equation (1) is suitable to describe the heat diffusion only in stationary conditions, while a Cattaneo-type equation, like Eq. (2), may be fruitfully used to describe the phenomenon also in the transient phase. The problem is then to provide a physical interpretation of the additional term introduced in Eq. (2). Cattaneo [1] was the first to interpret the delay time by introducing a sort of time-correlation of the particle collisions. Chester [15] pointed out that in non-metallic solids another microscopic process should be advocated, and he was the first to compare the theoretical forecasts of the heat waves velocity with that of collective elastic excitations, the phonons, getting a perfect match of the two.

Here in this section it is shown that the delay time introduced in the Cattaneo' equation may be physically interpreted as a relaxation time, which is a characteristic of the interaction processes in the Dual Model of Liquids.

### 2.1 The Dual Model of Liquids

The modelling of liquid structure has defied theorists for many years due to the difficulties of reproducing the experimental data. On the other hand, looking at the Clausius-Clapeyron plane, it is evident that the liquid state exists in an extremely limited range of pressure and temperature, if compared with the solid and gaseous states, making it very peculiar. In a series of recent papers [20-21] we have presented a model of organization of the liquid state at mesoscopic level, dubbed Dual Model of Liquids (DML), and applied to calculate several liquid specific quantities, the order-of-magnitude of the relaxation times governing the interaction process and, in particular, the thermal conductivity and the specific heat. The hypotheses behind the DML are actually two, and both have an experimental background. The first is that experiments conducted with the Inelastic X-ray Scattering and Inelastic Neutron Scattering techniques have made it possible to discover that the mesoscopic structure of liquids is characterized by the presence of solid-like structures, whose size is of a few molecular diameters, within which the elastic waves propagate as in the corresponding solid phase. The number and size of these structures varies with the temperature and pressure of the liquid. The second hypothesis is a direct consequence of the first. It consists in the assumption that elastic energy and momentum in liquids propagate by means of collective oscillations, or wave-packets, similar to phonons in crystalline solids. In this view, thermal energy is considered a form of elastic energy, as hypothesized by Debye [10-11], Brillouin [12-13] and Frenkel [26]. The liquid is then modeled at mesoscopic level in the DML as constituted by two sub-systems, mutually interacting among them, the wave-packets, or *lattice particles*, and the *liquid particles*, i.e. a sort of solid-like aggregates of liquids molecules. The two subsystems interact among them, exchanging energy and momentum, and making it possible to explain energy and mass diffusion in liquids. This picture recalls in some way that of Landau when he described the HeII as made of "normal" and "superfluid" parts [27]. One of the direct consequences of the above picture is that what we usually define "liquid" is not meant in the DML as a liquid following the classical definition, but a mixture of solid icebergs and an amorphous phase (i.e. a classical liquid). Of course, any liquid parameter whose magnitude is experimentally measured at temperatures and pressures where a usual liquid phase exists, is actually a pondered average of a solid/liquid value [20-21]. This concept will be useful and recalled later when we will introduce the several thermodynamic quantities needed to characterize the heat propagation in a liquid.

To avoid repeating concepts already described and extensively discussed elsewhere, we will gather here only the main aspects of the DML, inviting the readers who are interested to the details of the model to refer to the references [20-21] where the main experimental results, numerical simulations and theoretical developments are duly collected and deeply discussed. What matters here is shortly summarized below.

The internal energy $q_T$ per unit of volume of a liquid at temperature $T$ is:

3. $\quad q_T = \int\limits_0^T \rho C_V d\theta = f[\Theta / T]$

where $\rho$ is the medium density, $C_V$ the specific heat at constant volume per unit mass, $\Theta$ the Debye temperature of the liquid at temperature $T$. The fraction $q_T^{wp}$ of this energy:

4. $\quad q_T^{wp} = m q_T = \left\langle \mathcal{N}^{wp} \cdot \varepsilon^{wp} \right\rangle = m \dfrac{\int\limits_0^T \rho C_V d\theta}{\rho C_V T} \rho C_l T = m^* \rho C_l T$

is supposed be transported by the wave-packets. The parameter $m$ is the ratio between the number of collective DoF surviving at temperature $T$ and the total number of available collective DoF, then it holds $0 \le m \le 1$. $\mathcal{N}^{wp}$ is the number of wave-packets per unit of volume, and $\varepsilon^{wp}$ their average energy. $q_T^{wp}$ is propagated through the liquid by means of inelastic interactions between the *liquid particles* and the *lattice particles*, i.e. the wave-packets that transport the thermal, or generally the elastic energy. Figure 1 and Figure 2 show how the interaction between a *liquid particle* and a *lattice particle* works. To make more intuitive the model, we may ideally divide this elementary interaction in two parts: one in which the *lattice particle* collides with the *liquid particle* and transfers to it momentum and energy (both kinetic and potential) and one in which the *liquid particle* relaxes and the energy is returned to the thermal pool through a *lattice particle* alike in a tunnel effect. The energy $\Delta \varepsilon^{wp}$ and momentum $\Delta p^{wp}$ exchanged are respectively given by:

5. $\quad \Delta \varepsilon^{wp} = h \left\langle \nu_1 \right\rangle - h \left\langle \nu_2 \right\rangle = \Delta E_p^k + \Delta \Psi_p = -\nabla \phi^{th} \cdot \left\langle \Lambda_p \right\rangle$

6. $\quad \Delta p^{wp} = -\nabla \phi^{th} \cdot \left\langle \tau_p \right\rangle,$

where $\phi^{th}$ is the interaction potential between the *lattice particle* and the *liquid particle* that is in turn given by:

7. $\quad -\nabla \phi^{th} = f^{th} = \sigma_p \delta \left( \dfrac{J_q}{u_\varphi} \right)$

Some considerations regarding $f^{th}$, or $\phi^{th}$, are in order. First, $\phi^{th}$ is supposed to be anharmonic because of the inelastic character of the interaction, this last is in fact not instantaneous but lasts $\langle \tau_p \rangle$ during which the particle is displaced by $\langle \Lambda_p \rangle$ (here and in the rest of the paper, the two brackets $\langle \ \rangle$ indicate the average over a statistical ensemble of the quantity inside them). Second, $f^{th}$ can be positive or negative depending on whether the quantity $\left( J_q / u_\phi \right)$ increases or decreases as consequence of the interaction. In equations (5)-to-(7), $\langle \nu_1 \rangle$ and $\langle \nu_2 \rangle$ are the wave-packet (central) frequency before and after the interaction, $\Delta E_p^k$ and $\Delta \Psi_p$ are kinetic and potential energy acquired by the liquid particle as consequence of the interaction. Finally, $u_\phi$ is the phase velocity associated with the wave-packet and $\sigma_p$ is the cross-section of the "obstacle", the solid-like cluster, on the surface of which $f^{th}$ is applied. The reader who wants to learn more about the above is warmly addressed to the related literature [20-21].

The event of Figure 1a, in which an energetic *lattice particle* transfers energy and momentum to a *liquid particle*, is commuted upon time reversal into the one of Figure 1b, where a *liquid particle* transfers energy and momentum to a *lattice particle*. The *liquid particle* changes velocity and the frequency of *lattice particle* is shifted by the amount $\left( \langle \nu_2 \rangle - \langle \nu_1 \rangle \right)$. Due to its time symmetry, we have assumed this mechanism be the equivalent of Onsager' reciprocity law at microscopic level [8-9,20]. In a pure isothermal liquid energy and momentum exchanged among the icebergs are statistically equivalent, and no net effects are produced. Said in other words, events of type a) are equally likely to occur as events of type b) and will alternate among them to keep the balance of the two energy pools unaltered. Besides, the macroscopic equilibrium will ensure also the mesoscopic equilibrium; events a) and b) will be equally probable along any direction, to have a zero average over time and space. On the contrary, if a symmetry breaking is introduced, as for instance in the case of a temperature gradient which we are going to discuss in this paper, events of type a) will statistically prevail over events of type b) along the preferential direction of the externally applied temperature gradient.

Figure 2 is a close-up of the first part of the "*wave-packet* $\leftrightarrow$ *liquid particle*" interaction shown in Figure 1a, during which the phonon transfers energy and momentum to the liquid particle. $\langle \Lambda_{wp} \rangle$ is the extension of the wave-packet and $\langle d_p \rangle$ that of the liquid particle. Once $\langle \tau_p \rangle$ has elapsed and the liquid particle has travelled by $\langle \Lambda_p \rangle$, the particle relaxes the energy stored into internal DoF; then it travels by $\langle \Lambda_R \rangle$ during $\langle \tau_R \rangle$ (not shown in the figure) [20].

The presence of relaxation time is a signature of liquids. Although the thermal unrest in liquids is similar as in solids, the positions of atoms in liquids are temporary rather than permanent.

After having performed a number of oscillations around a given position, the atom in a liquid jumps to another equilibrium position. This step-by-step wandering leads to a gradual mixing up of the atoms [see 20, in particular Figures 1 and 2 therein]; in liquids it proceeds much faster than in solids, and has a simpler character because of the absence of definite lattice sites.

From the above and from Figure 1, the physical meaning and origin of the relaxation time $\langle \tau \rangle$ present in the Cattaneo equation jump clearly to the eye: $\langle \tau \rangle$ is the time interval during which the energy subtracted from the wave-packet's current, i.e. from the thermal energy current, is sequestered into a non propagative form, that is the potential energy $\Delta \Psi_p$ (Eq. (5)) pertaining to the *liquid particle* internal DoF. This energy content is then given back to the heat current a $\langle \Lambda \rangle$ step forward, alike in a tunnel effect. This is the physical mechanism in the DML constituting the relaxation effect of heat propagation during the transient phase, i.e. when the Cattaneo equation correctly describes the time evolution of the thermal energy propagation (or of the temperature distribution) in a system out of equilibrium. Similar reasoning can be made for a flux of liquid particles when a concentration gradient is imposed to the system, it is indeed enough to follow the reverse time evolution of the elementary interaction between a *liquid particle* and a *lattice particle*, illustrated in Figure 1b.

What happens in a liquid at mesoscopic level when a temperature gradient is externally applied to the system? In the previous paper [20] we have deeply analyzed the Stationary State (SS), i.e. when every transient effect is overcome and a stabilized and uniform temperature gradient has been reached. Here we will concentrate our attention on the Transient Phase (TP), i.e. the time interval during which the thermal front goes through the system.

Let $\nabla T^{TP}$ be the temperature gradient that is forming across the system. To fix the ideas, let $T_h$ be the temperature of the heat source applied to the system. Because the thermal front travels through the system, it will be

8. $\qquad \nabla T^{TP}(z) = \dfrac{T_h - T_c(z)}{\Delta z(t)}$,

where $T_c(z)$ is the temperature of the advancing thermal front at the point $z$, and $z$ the coordinate of the advancing thermal front at time $t$. Due to the application of the external temperature gradient, the heat flux crossing the system will give rise to an increase of the number of *wave-packet $\leftrightarrow$ liquid particle* collisions in the direction of the heat flux $J_q$. In fact, in this case, events of type a) in Figure 1 have a larger probability to occur than events of type b). At equilibrium the flux of wave-packet is driven in the DML by the presence of a *virtual temperature gradient* [20], $\left\langle \dfrac{\delta T}{\delta z} \right\rangle$.

If $\langle v_p \rangle$ is the average number per second of *wave-packet ↔ liquid particle* collisions due to $\left\langle \dfrac{\delta T}{\delta z} \right\rangle$, along the direction of $J_q$ such number is $\langle v_p \rangle / 6$ for symmetry reasons. When $\nabla T^{TP}$ is applied, there will be an imbalance in such number, precisely an increase of collisions along $z$; let $\delta \langle v_p \rangle$ indicate the increase in the number of *wave-packet ↔ liquid particle* collisions per second due to $\nabla T^{TP}$. $\delta \langle v_p \rangle$ is a quantity neither constant in time nor in space, because such is $\nabla T^{TP}$ given by Eq. (8). When the SS is reached the temperature gradient and the heat flux will have reached their stationary values, $\delta \langle v_p \rangle$ even will reach its stationary value (that used in [20]).

To evaluate $\delta \langle v_p \rangle$ we suppose as first approximation that the application of $\nabla T^{TP}$ increases proportionally the number of *wave-packet ↔ liquid particle* collisions:

9.     $\dfrac{\langle \delta v_p \rangle}{\langle v_p \rangle / 6} = \dfrac{\nabla T^{TP}(z)}{\delta T / \delta z}$

Consequently, every *liquid particle* will execute as many jumps per second in excess as $\delta \langle v_p \rangle$ along the direction of $J_q$, each of average length $\langle \Lambda \rangle$, for a total distance travelled $\langle \Lambda \rangle \delta \langle v_p \rangle$ per second along the direction of heat flux generated by the external local temperature gradient. This quantity represents the drift velocity $\left\langle v_p^{th} \right\rangle^{TP}$ of the *liquid particle* during the *TP* along $z$ due to the external temperature gradient:

10.     $\left\langle v_p^{th} \right\rangle^{TP} = \langle \Lambda \rangle \langle \delta v_p \rangle = \dfrac{\langle \Lambda \rangle \langle v_p \rangle}{6} \dfrac{\nabla T^{TP}(z)}{\langle \delta T / \delta z \rangle}$.

The reader must not be fooled by the similarity of Eq. (10) with the analogous expression given in [20] for the *SS* (see Eq. 46 in reference 20), from which one could hastily and erroneously conclude that in the *SS* speed and imbalance of impacts are greater than in the *TP*. In fact, the value of $\left\langle v_p^{th} \right\rangle^{TP}$ in (10) is a function of $\nabla T^{TP}$, which in the *TP*, as is known, can have much greater values than that of the stationary state, being it applied over much shorter distances. At steady state $\left\langle v_p^{th} \right\rangle^{TP}$ will assume the value $\left\langle v_p^{th} \right\rangle$ given in [20].

In the previous reasoning we assumed that the heat flux inside the system is due to the presence of an external heat source at a temperature $T_h$ higher than that of the system. The principle on which the reasoning is based, however, also holds true in the case in which the flow of heat inside the system is due to the presence of a heat sink at a temperature $T_c$ lower than that of the system, or even in the case in which the system is in contact with two different sources, one with a temperature higher than that of the system and the other with a lower, keeping the average temperature of the system unchanged. Of course, if there will always be an imbalance in the average number of collisions per second in the direction of the heat flow, the previous situations will obviously differ from each other due to the different final thermal content of the system.

With reference to Figure 1 and Figure 2, the global effect of the interactions is to transfer energy from one point of the liquid to another, exploiting the excitement of the internal DoF of the icebergs, which have characteristic structural relaxation times. When the wave-packet hits a liquid particle the force $f^{th}$ develops, it acts for $\langle \tau_p \rangle$ seconds displacing the particle by $\langle \Lambda_p \rangle$. Energy $\langle \Delta\varepsilon^{wp} \rangle$ and momentum $\langle \Delta p^{wp} \rangle$ given by Eqs.(5) and (6) respectively are transferred to the particle during $\langle \tau_p \rangle$, increasing its kinetic and potential energy and exciting internal vibrational energy levels. These DoF oscillate similarly to those pertaining to solid state, giving origin to (quasi) elastic waves with characteristic length $\langle \Lambda_0 \rangle$ and period $\langle \tau_0 \rangle$. $\langle \tau_R \rangle$ seconds later and $\langle \Lambda_R \rangle$ meters forward they are given back to the wave-packet pool, and the process is repeated again $\langle \tau_{wp} \rangle$ seconds later and $\langle \Lambda_{wp} \rangle$ meters forward, $\langle \tau_{wp} \rangle$ being the phonon mean free flight time and $\langle \Lambda_{wp} \rangle$ their mean free path, i.e. the distance travelled between two interactions with *liquid particles*. What matters here to understand the dynamics linked to the relaxation time, is to recognize that $\langle \tau \rangle$ is the time interval during which the energy disappears from the liquid thermal pool because it is trapped in the internal DoF; once $\langle \tau \rangle$ has elapsed, it reappears in a different place. In this way relaxation times, introduced *ad hoc* by Frenkel, find an immediate physical interpretation. The wave-packet emerges from the collision with reduced energy and momentum, while the particle acquires the energy and momentum lost by the phonon. The effect is that of having converted part of the energy carried by the phonon into potential energy of the liquid particle.

The interaction mechanism described above not only foresees the transport of energy, but also that of mass, which are always coupled in such a system; consequently it may be fruitfully exploited even to model the mutual diffusion of a solute with respect to the solvent in a liquid system submitted to a temperature gradient (Soret effect) or to a concentration gradient (Dufour effect). These topics however will be dealt with in separate papers [28].

Of course, as deeply discussed already in [20], such interactions, and consequently the relaxation mechanism, are always operating in a liquid whatever its thermodynamic state, therefore they are present not only during the *TP*, but also when the system is in thermal equilibrium or with an external temperature gradient at the *SS*. The distinguishing aspects between the two cases consist in the fact that i) when the system is at thermal equilibrium, these interactions are equally likely to occur along any direction, thus they do not have any effect at macroscopic level; ii) when the system reaches the stationary equilibrium, i.e. when the temperature does not vary anymore vs time, their effect is equally not manifesting at macroscopic level because it is present everywhere in the system, as the thermal current has established and stabilized along the entire system. Said in other words, every transient effect has been overcome in the dynamics of the system at the microscopic level. This implies that the time elapsed from the beginning of the heat propagation has become large enough to make negligible the additional term $\tau \dfrac{\partial J_q}{\partial t}$ characterizing the Cattaneo equation (2), that reduces now to the classical Fourier law (1).

In the next section we will show how a hyperbolic equation tailored on the wave-packet current is the one suitable for describing the flow of heat (or mass) in a Dual System.

## 2.2 The heat propagation equation "with memory" in the DML.

The duality of liquids in the frame of the DML defined above, i.e. constituted by two interacting subsystems, allows to identify the physical mechanism – the elementary interaction between *liquid particles* and *lattice particles* – by means of which the two subsystems interact, and consequently also the relaxation time $\langle \tau \rangle$ at the base of the "memory term" that is responsible of the modification of the Fourier law (Eq.1) into a Cattaneo equation (Eq.2). As consequence of the interaction, the energy carried by wave-packets is temporarily stored into a non propagating form, the potential energy of the internal DoF of the liquid particle. Once the relaxation time $\langle \tau \rangle$ has elapsed, that energy "get out from the tunnel" and is given back to the thermal pool.

If Cattaneo [1] was the first to identify on a physical basis the need to modify the Fourier law by introducing a relaxation term to correctly describe the initial phases of the propagation of the thermal signal, the same type of equation had already been proposed years earlier by Frenkel [26] as basic equation of Maxwell's elasticity relaxation theory [2]. Maxwell indeed identified in his model only the need to introduce a relaxation term, but he did not get to write the relative equation. Even more curiously, Frenkel arrived at the formulation of the equation (although formulated in terms of Navier-Stokes equation), but did not solve it. In the DML model, we start exactly where Frenkel stopped, however making appropriate tailoring of the equation itself to adapt it to the phenomenon it describes. Frenkel's starting point was the same as that of Maxwell, namely the idea that in liquids elastic responses, typical of solids, and viscous, typical of fluids, are combined; in one word, liquids exhibit *viscoelastic* responses. The prevalence of one of the two responses over the other is due to the conditions in which the liquid lies, to the type of stimulus to which it is subjected, and, finally, to the scale on which the stimulus is studied, i.e. macroscopic or mesoscopic. Let's then start to examine the consequences of our approach in the propagation of energy into liquids by re-writing the Cattaneo equation (2) in the slightly modified form

11. $\quad J_q^{wp} + \langle \vartheta \rangle \dfrac{\partial J_q^{wp}}{\partial t} = -K_l^{wp} \dfrac{\partial T}{\partial z}$

in which we used the terms related to the collective excitations, with $\langle \vartheta \rangle = n \langle \tau \rangle$, $n \geq 1$ (for reasons which will be clarified later on). The quantity $\dfrac{\partial J_q^{wp}}{\partial t}$ is intrinsically negative because it represents the amount of energy subtracted to the heat flux and transformed into kinetic and potential energy of the icebergs. The relaxation time $\langle \tau \rangle$ has been evaluated [20] amounting to several picoseconds, consequently $\langle \vartheta \rangle$ is of the same order of magnitude, or a little bit larger. We will return in the **Discussion** on the meaning of $\langle \vartheta \rangle$; by now it is enough to point out that it equals several $\langle \tau \rangle$'s. As for the thermal conductivity, we use the expression derived for the wave-packets [20]:

12. $\quad K_l^{wp} = \dfrac{1}{3} u^{wp} \langle \Lambda_{wp} \rangle \dfrac{\partial}{\partial T} \left[ \langle \mathcal{N}^{wp} \varepsilon^{wp} \rangle \right]_V = \dfrac{1}{3} u^{wp} \langle \Lambda_{wp} \rangle \rho C_V^{wp} = \dfrac{1}{3} u^{wp} \langle \Lambda_{wp} \rangle m \rho C_V \left[ \dfrac{m^*}{m^2} \dfrac{dm}{dT} T + 1 \right]$

where $u^{wp} = \dfrac{\langle \Lambda_{wp} \rangle}{\langle \tau_{wp} \rangle} = \lambda^{wp} \cdot \nu^{wp}$ is the wave-packet velocity, and the last equality is obtained by

means of the wave-packet' specific heat at constant volume $C_V^{wp}$ [20-21]

13. $\quad \begin{cases} \rho C_V^{wp} = \dfrac{\partial q_T^{wp}}{\partial T} \bigg|_V = \dfrac{\partial}{\partial T} (m q_T)_V = \left( q_T \dfrac{dm}{dT} + m \dfrac{\partial q_T}{\partial T} \right)_V = q_T \dfrac{dm}{dT} + m C_V = \\ \quad\quad = \rho m C_V \left[ \dfrac{q_T}{m \rho C_V} \dfrac{dm}{dT} + 1 \right] = m \rho C_V \left[ \dfrac{m^*}{m^2} \dfrac{dm}{dT} T + 1 \right] \end{cases}$

with $C_V^{wp} \leq C_V$. Combining Eq.(11) with the continuity equation, with $C_p^{wp}$ being the isobaric heat capacity [2],

14. $\quad \rho C_p^{wp} \dfrac{\partial T}{\partial t} + \dfrac{\partial J_q^{wp}}{\partial z} = 0$

we finally get the hyperbolic propagation equation in terms of temperature $T$:

15. $\quad \dfrac{\partial^2 T}{\partial z^2} = \dfrac{\rho C_p^{wp}}{K_l^{wp}} \dfrac{\partial T}{\partial t} + \langle \vartheta \rangle \dfrac{\rho C_p^{wp}}{K_l^{wp}} \dfrac{\partial^2 T}{\partial t^2}$

Of course, an analogous expression may be obtained for the heat flow $J_q^{wp}$ (or for a matter flow, in case one would consider the mass propagation instead of heat).

Equations like (15) are encountered in physics in the several areas where two subsystems belonging to the same closed system interacts among them. As is known from mathematics it yields complex solutions (frequencies), but only those with non-zero real parts are representative of propagating modes, of interest here in this paper. Customizing the method adopted in [22-25, 29], let's then suppose that a general solution of Eq. (15) is a plane wave with equation:

16. $\quad T(z,t) = \widetilde{T} e^{i(kz - \omega t)}$

By replacing Eq.(16) into Eq.(15) we get the associated algebraic equation:

17. $\quad \omega^2 + \dfrac{i}{\langle \vartheta \rangle} \omega - \upsilon_C^{wp\,2} k^2 = 0$

with

18. $\quad \upsilon_C^{wp} = \sqrt{\dfrac{K_l^{wp}}{\rho C_p^{wp} \langle \vartheta \rangle}} = \sqrt{\dfrac{D_l^{wp}}{\langle \vartheta \rangle}}$

where $D_l^{wp}$ is the liquid thermal diffusivity. An interesting alternative expression for $\upsilon_C^{wp}$ may be obtained by replacing in eq.(18) the expression for $K_l^{wp}$ from eq.(12), $\upsilon_C^{wp} = \sqrt{\dfrac{u^{wp} \langle \Lambda_{wp} \rangle}{3\gamma^{wp} \langle \vartheta \rangle}}$ with

$\gamma^{wp} = C_p^{wp} / C_V^{wp}$. It shows that $\upsilon_C^{wp}$ depends either on the relaxation time and on the flight

---

[2] So far we have not derived yet in the DML an expression for the specific heat at constant pressure, $C_p^{wp}$, however we expect the same relation holds as for the classical thermodynamic variables, i.e. that $C_p^{wp} > C_V^{wp}$ because $C_p^{wp}$ accounts for the enthalpy $h^{wp}$ variation vs temperature rather than the internal energy $q_T^{wp}$.

parameters of the wave-packets, as one would have expected. Eq. (17) yields complex solutions (frequencies), given by:

19. $\quad \omega = -\frac{i}{2\langle\vartheta\rangle} \pm \sqrt{\left(\upsilon_C^{wp}k\right)^2 - \frac{1}{4\langle\vartheta\rangle^2}} = -\frac{i}{2\langle\vartheta\rangle} \pm \omega_C$ .

For Eq.(17) to have solutions with non-zero real parts, corresponding to propagative waves, it must hold:

20. $\quad k > \frac{1}{2}\sqrt{\dfrac{\rho C_p^{wp}}{K_l^{wp}\langle\vartheta\rangle}} = \frac{1}{2}\sqrt{\dfrac{1}{D_l^{wp}\langle\vartheta\rangle}} = k_m$

or alternatively:

21. $\quad \langle\vartheta\rangle < \dfrac{\rho C_p^{wp}}{4K_l^{wp}k_m^2} = \dfrac{1}{4D_l^{wp}k_m^2} = \langle\vartheta\rangle_M$ .

The final solution of Eq.(15) is then:

22. $\quad T(z,t) = T_0 e^{-(1/2\langle\vartheta\rangle)t} e^{i(kz\pm\omega_C t)}$

In particular, Eq.(20) tells us that the dynamics described above works above a minimum value for the wave vector, $k_m$. Said in other words, heat propagation is inhibited for waves with wave vector below $k_m$. This means that we are in presence of Gapped Momentum State (GMS), or simply the $k$-gap, that is always present in all systems where the propagation of energy is described by equations such as Eq.(15), or where the dispersion relation is like Eq.(17) [22-25]. Eq.(21) provides a way to calculate $\langle\vartheta\rangle_M$, the maximum relaxation time, once $k_m$ is known from experiments. Finally, differently from the Fourier equation, Eq. (15) contains also the second time derivative of $T$ multiplied by the relaxation time $\langle\vartheta\rangle$, thus allowing for damped propagative waves as solutions, with damping constant $\langle\vartheta\rangle$ and velocity $\upsilon_C^{wp}$ given by eq.(18). We will deeply analyze the consequences of Eqs. (15), (20) and (21) in the **Discussion**.

## 3. Discussion

Let us first point out that, as is well known, Eq.(15) is a truncated form of a more extensive expression which includes pressure and density spatial variations of the several quantities present in it, and that may be derived from kinetic theory in the case of an ideal gas [2,3,14,19]. Besides, many additional terms are present in the complete form if one considers the spatial variations of mechanical properties of the medium. These corrections however are not germane to the matters which follow, so they have been ignored as specified in the **Introduction** (they should be taken into account in a careful treatment dealing with a practical experimental situation).

Equation (15) is the simplest equation combining diffusion and waves, i.e. propagation. Besides, and this is not secondary, it allows for dissipation to take place. Its meaning is well known: it states that there is a physical mechanism by means of which energy is not freely flowing through the liquid, but is temporarily subtracted for a time $\langle \vartheta \rangle$; this effect has an influence during the transient phase but has not at the steady state (although still present at microscopic level), because it becomes irrelevant when a time interval larger that $\langle \vartheta \rangle$ has elapsed. Hence $\langle \vartheta \rangle$ represents the finite build-up time for the onset of a thermal current once a temperature gradient is applied to a system. The heat flow does not start instantaneously but rather grows gradually with the delay time $\langle \vartheta \rangle$. Conversely, if a thermal gradient is suddenly removed, there is a lag in the disappearance of the heat current and Eq.(15) exhibits just such a delay, whereas the classical Fourier diffusive equation does not [29]. At the best of the author' knowledge, this is the first time the telegrapher equation is used to describe the propagation of phonons in normal liquids as carriers of the thermal (and elastic) signal[3]. The presence of the two interacting populations, *lattice particles* and *liquid particles*, justifies the presence of the relaxation term in the equation for heat transport (11), that we have even proposed to be linear in the relaxation time. These interactions are characterized by a tunnel effect which is the physical mechanism allowing the interpretation of the heat transport to be of a dynamical nature, thus avoiding the dilemma of infinite diffusion velocity.

Remaining on eq.(15), it is interesting to note that the quantity $\left( \langle \vartheta \rangle \dfrac{\partial J_q^{wp}}{\partial t} \right)$ is the fraction of the heat flux carried by wave-packets transformed into energy of molecular modes of the *liquid particles*; dividing it by $u^{wp}$, we get the momentum flux $J_p^{lp}$ transferred from the heat current to the *liquid particles*:

23. $\quad J_p^{lp} = \dfrac{1}{u^{wp}} \left( \langle \vartheta \rangle \dfrac{\partial J_q^{wp}}{\partial t} \right)$

---

[3] However, by digging into the depths of literature, one can discover how already in 1946 Peshkov [63] had hypothesized that in low temperature liquids "*a gas of thermal quanta capable of performing vibrations similar to those of sound should exist*".

corresponding to the kinetic energy density of *liquid particles*. Dimensionally $J_p^{lp}$ is also a pressure, and as such it represents the radiation pressure exerted by the current of wave-packets on the *liquid particles* following the interaction. This pressure may also be regarded as a sort of osmotic pressure determining the displacement of *liquid particles* following the collisions with the wave-packets. Indeed, to a pressure difference $\Delta P^{wp} \equiv J_p^{lp}$ the liquid responds with the self-diffusion of the *liquid particles*. This interpretation could be used to explain the selective diffusion of solute (or solvent) in a solution, either in thermal equilibrium and in presence of a gradient. The tunnel effect could be regarded as a semi-permeable membrane, that works allowing only the passage of *liquid particles* and preventing that of *lattice particles* [28]. A similar argument was raised by Ward and Wilks [49] in dealing with the machano-thermal effect observed in HeII.

The delay time $\langle \vartheta \rangle$ is associated with the "communication time" between phonons and molecule' clusters for the commencement of resistive flow. Fourier equation may be considered indeed as the thermal counterpart (at steady state) of the first Ohm's law, the flow of electrical charges being replaced by the resistive flow of thermal charges, the phonons. If, on one hand, $\langle \vartheta \rangle$ is the time necessary for the establishment of the resistive flow, $\Phi = 1/\langle \vartheta \rangle$ is proportional to the frequency of occurrence of the *phonon* $\leftrightarrow$ *particle* collisions. $\Phi$ is therefore a measure of how quickly the phonons are able to transmit the thermal signal in the liquid as a result of their interactions with the clusters. It is intriguing to compare eq.(15) with that obtained by Frenkel upon the generalization of the Navier-Stokes equation to viscoelastic media [22-26], when he considered the Maxwell relaxation theory of elasticity, namely:

24. $\quad \dfrac{\partial^2 u}{\partial z^2} = \dfrac{\rho}{\eta_l}\dfrac{\partial u}{\partial t} + \tau \dfrac{\rho}{\eta_l}\dfrac{\partial^2 u}{\partial t^2}$

where $u$ is the fluid velocity component normal to $z$ and $\eta_l$ its viscosity. Such a comparison shows that the ratio $\eta^{wp} = \dfrac{K_l^{wp}}{C_p^{wp}}$, dimensionally a viscosity, has a similar role for the temperature as $\eta_l$ for the fluid current. Therefore, we could speculate that if $\eta_l$ represents the capability of a fluid to transmit momentum, $\eta^{wp}$ represents the same capability for the wave-packet current. Indeed, bartering eq.(13) for $C_p^{wp}$, the ratio $\dfrac{K_l^{wp}}{C_p^{wp}}$ is just the momentum per unit of surface carried by wave-packets.

Very instructive is however the comparison of eq.(15) with the analogous in electromagnetism represented by the telegraphy equation. The comparison with electromagnetism is indeed the most suitable because of the presence of flowing electrical charges, as the wave-packets in the DML. Let's indeed suppose to have a circuitry in which $L$ represents the uniformly distributed inductance of the line, $I$ is the electrical current flowing through the line, $V$ the electrical potential and $R$ the uniformly distributed electrical resistance of the line. By excluding sources and sinks along the line, the equation that relates the voltage to the current in terms of properties per unit length of the conductor is:

25. $\quad -\dfrac{\partial V}{\partial z} = IR + L\dfrac{\partial I}{\partial t}$

and the charge conservation equation is:

26. $\quad -\dfrac{\partial I}{\partial z} = C\dfrac{\partial V}{\partial t}$

where $C$ is the uniformly distributed capacity of the line. Combining Eqs.(25) and (26)

27. $\quad \dfrac{\partial^2 V}{\partial z^2} = RC\dfrac{\partial V}{\partial t} + LC\dfrac{\partial^2 V}{\partial t^2}$

is the well-known equation of telegraphy where we easily recognize eq.(15) by replacing $V$, $R$, $L/R$ and $C$ with $T$, $1/K_l^{wp}$, $\langle \vartheta \rangle$ and $\rho C_p^{wp}$, respectively. Similarly, we may recognize eq.(11) by replacing even $I$ with $J_q^{wp}$ in eq.(25). We may conclude that considering the heat in a liquid being carried by wave-packets current is equivalent considering an electric line affected by inductance.

An interesting aspect of Eq.(15) is its behaviour in the high frequency limit of fast thermal fluctuations. To fix the ideas, let us suppose that the external temperature $T$ varies at a rate $f = \dfrac{1}{T}\dfrac{\partial T}{\partial t}$, much faster than $\Phi = 1/\langle \vartheta \rangle$ introduced before, then Eq.(15) predicts a wave propagation of temperature instead of diffusion. The frequency $\Phi$ is also a critical frequency for the onset of thermal waves; if the external temperature varies with time at a rate higher than $\Phi$, the temperature signal has no way to stabilize, and the diffusive regime does not take place. This frequency is directly proportional to the thermal resistivity, being zero if the thermal resistivity is zero. The order of magnitude of the relaxation time has been calculated [20] amounting to several picoseconds, in agreement with experiments; consequently, $\Phi \approx 10^{11} \div 10^{12}\,Hz$.

Following the previous reasoning we could affirm that the Eq.(15) takes into account also for the capability of a medium to dissipate the thermal energy carried by the (anharmonic) wave-packets, a circumstance not considered by the classical Fourier equation, that is formally the extension of Eq.(15) to the steady-state. In fact, neglecting dissipation in Eq.(15) corresponds to consider the limit $\langle \vartheta \rangle \to 0$. From the physical point of view, a finite value of $\langle \vartheta \rangle$ means that the thermal wave has a finite propagation range, and here we come to comment Eqs. (17), (20) and (21) more in detail. The first observation is related to Eq.(17) that represents a dispersion relation for the wave-packet' frequency (or at least for those interacting with the *liquid particles*.) It represents a fingerprint of Dual Systems, i.e. of closed systems made of two distinct and interacting populations. In fact, as indicated from Eq.(20), not all wave-packets are allowed to interact but only those with momentum above $k_m$, revealing the so-called *k*-gap. This aspect has been extensively discussed elsewhere [20,22-25], here we want however to put the attention also on Eq.(21) and on the following expression, that provides the gap in terms of wavelength instead of momentum:

28.   $\langle \Lambda_C \rangle < \sqrt{\dfrac{K_l^{wp} \langle \vartheta \rangle}{\rho C_p^{wp}}} = \sqrt{D^{wp} \langle \vartheta \rangle} \Rightarrow \langle \Lambda_C \rangle_M = \sqrt{D^{wp} \langle \vartheta \rangle_M}$ .

The reader has certainly recognized the Einstein relation [17] in the last equality. The same expression for $\langle \Lambda_C \rangle$ as Eq.(28) is obtained by observing that it holds

29.   $\langle \Lambda_C \rangle = \upsilon_C^{wp} \langle \vartheta \rangle$ .

As before, using eq.(12) into eq.(28), one gets an expression for $\langle \Lambda_C \rangle_M$ in terms of relaxation time and wave-packet' parameters, $\langle \Lambda_C \rangle_M = \langle \Lambda_{wp} \rangle \sqrt{\dfrac{\langle \vartheta \rangle_M}{3\gamma^{wp} \langle \tau_{wp} \rangle}}$ .

In discussing the solutions of Eq. (24) in [24] the authors have rightly wondered how large can be the *k*-gap be. Here we move the attention on the $\langle \Lambda_C \rangle$-gap, or on the $\langle \vartheta \rangle$-gap, rather than on the *k*-gap. Recalling that a wave is well defined only if its wavelength is smaller than the propagation distance, eqs.(21) and (28) tell us that there is an upper limit also for the relaxation time and for the distance travelled by wave-packets during the propagation of the thermal signal or, what is the same, for the maximum length of the tunnel or alternatively for the maximum number of *liquid particles* with which a *lattice particle* may interact before being definitively damped (remember that in Eq. (11) it holds $\langle \vartheta \rangle = n \langle \tau \rangle$ with $n \geq 1$). This gives us the possibility to illustrate the dynamics occurring in a liquid during the thermal transient. When an external temperature gradient is applied, the thermal content given by eq.(4) increases, determining an imbalance of the phonon flux, so that there will be an excess of interactions as well of the energy transferred from the thermal current to the *liquid particles* through events like those of Figure 1a (see eq.(9)). Let us start at $z = z_0$. Each interaction lasts $\langle \tau \rangle = \langle \tau_p \rangle + \langle \tau_R \rangle$, the time interval during which the energy disappears as liquid free energy to become iceberg' internal and kinetic energy, and the *liquid particle* moves by $\langle \Lambda \rangle = \langle \Lambda_p \rangle + \langle \Lambda_R \rangle$ forward, after that the emerging wave packet has lost part of its initial polarization. Depending on its residual energy and momentum, it may interact with another liquid particle $\langle \tau_{wp} \rangle$ seconds later, and the above process is replicated, say, *n* times. The overall duration of the randomisation process lasts $\langle \vartheta \rangle = n \langle \tau \rangle$, during which the particle, and the thermometric front, will have advanced by $\langle \Lambda \rangle_C = v_C^{wp} \cdot n \cdot \langle \tau \rangle$, determining a temperature increase by $\Delta T$ over $\langle \Lambda_C \rangle$. At the end of $\langle \vartheta \rangle$ a liquid warmer both in its molecular and gas of excitations components will be in contact with the still unperturbed medium laying beyond $z = z_0 + \langle \Lambda_C \rangle$, setting the stage for a replica of the events. At $z = z_0$ instead, in absence of a new advancing front, the process of heat transport after $\langle \vartheta \rangle$ seconds has reached the steady-state, the propagation of thermal excitations becoming at this point purely diffusive.

Always in [24] the authors highlight how the ***k***-gap sets three distinct intervals for the propagating modes, namely: i) non propagating shear modes, ii) damped oscillatory shear modes and iii) purely elastic non dissipative shear modes. The authors recognize the presence of solid-like structures in liquids for distances where purely elastic non dissipative modes are allowed, a picture absolutely similar to that of the DML. Comparing then $\langle \vartheta \rangle_M$ and $\langle \Lambda_C \rangle_M$ with $\langle \tau \rangle$ and $\langle \Lambda \rangle$ of Eqs. (5) and (6), respectively (or with their multiples, $\langle \vartheta \rangle = n \langle \tau \rangle$ and $\langle \Lambda_C \rangle = n \langle \Lambda \rangle$, with $n \geq 1$), we may envisage the following correspondences: i) the momentum carried by wave-packets is too much low to interact with *liquid particles*, and there are no propagating modes ($n = 0$); ii) the momentum is large enough to allow the wave packet to interact with several *liquid particles*, thus getting damped oscillating modes and $n > 1$; iii) this is what happens inside a single *liquid particle*, being a solid-like structure and the propagation of elastic perturbations dispersionless. In this last case one experiences the propagation velocity of elastic waves similar to that of the corresponding solid [20,30-44]. These limitations are therefore an indirect evidence of the presence of pseudo-crystalline structures in liquids.

It is interesting, if not mandatory, to numerically evaluate the values of phonon mean-free-path $\langle \Lambda_0 \rangle$ and life-time $\langle \tau_0 \rangle$ [20] within a liquid particle, see Figure 2, that are connected with $\langle \Lambda \rangle_C$ and $\langle \vartheta \rangle$. They can be deduced from the experimental values obtained in light scattering experiments for the wave-vector $k$. Indeed, $\langle \Lambda_0 \rangle$ will be a multiple of the phonon wavelength $\lambda^0$, $\langle \Lambda_0 \rangle = a\lambda^0$, and $\langle \tau_0 \rangle$ of $\tau = 1/\nu^0$, $\langle \tau_0 \rangle = a/\nu^0$, with $a > 1$. Using the data for water of [42,44], typical values for the parameters characterizing a phonon (variation range is function of temperature, pressure and $k$ orientation) are: (central) frequency $\dfrac{n}{\langle \tau_0 \rangle} = \langle \nu^0 \rangle \approx 0,95 \div 2,5\ THz$, wave-length $\dfrac{\langle \Lambda_0 \rangle}{n} = \langle \lambda^0 \rangle \approx 1 \div 3\ nm$ and velocity $\dfrac{\langle \Lambda_0 \rangle}{\langle \tau_0 \rangle} = \langle \lambda^0 \rangle \cdot \langle \nu^0 \rangle = \langle u^0 \rangle \approx 3.1 \div 3.4 \cdot 10^3\ m/s$. Interestingly, this last value is in very good agreement with the experimental data obtained for the propagation velocity of thermal waves in water [30], of $3.2 \cdot 10^3\ m/s$ (see also Figure 5 in [20]).

It is worth noting that the finite extension of the $\langle \Lambda_C \rangle$-gap is a direct consequence of the presence of dissipation, that takes place during $\langle \vartheta \rangle$. The absence of dissipation would imply an infinite propagation range, as it happens in perfect ideal crystals and as is described by the classical Fourier law. Because of the relevance of these aspects, they will be carefully dealt with in a separate paper [28], being outside the main topic of the present one. However it is worth anticipating some additional comments regarding the presence of **k**-gap [22-25]. Because the **k**-gap in liquids is present only in the transverse spectrum, while the longitudinal one remains gapless, it is reasonable assuming the **k**-gap in liquids be related to a finite propagation length of shear waves. For the **k**-gap to emerge in the wave spectrum two essential ingredients are mandatory. The first is to get a wave-like component enabling wave propagation; this is represented in the DML by the wave packets, that in turn allow for propagation of the thermal (and generally of the elastic) signal through progressive waves. The second consists in getting a dissipative effect that disrupts the wave continuity and dissipates it over a given distance, thus destroying waves and giving origin to the **k**-gap. This latter is represented by the wave-packet $\leftrightarrow$ *liquid particle* interaction, which works by moving the wave-packets from where they are absorbed by the *liquid particle*, to where they return to the system' energy pool, alike in a tunnel effect. With $\langle \vartheta \rangle$ the time during which the shear stress relaxes, then $\langle \Lambda \rangle_C = \upsilon_C^{wp} \cdot \langle \vartheta \rangle$ gives the shear wave propagation length (or liquid elasticity length).

Which is the physical interpretation of the behaviour described above in the frame of DML? The additional time derivative of the flow of wave packets, $\dfrac{\partial J_q^{wp}}{\partial t}$, is intrinsically negative because it represents that part of the energy flow that is subtracted during the phonon propagation and sequestered into internal DoF of the icebergs. We may imagine that thermal (and elastic) energy does not propagate continuously, but step-by-step, as originally supposed by Frenkel [see also Figs.(1) and (2) in ref.20]. The energy remains "trapped" in the internal DoF for a certain time $\langle \tau \rangle$, at the end of which the particle relaxes the stored energy, subtracted to the pool a mean free path $\langle \Lambda \rangle$ before. This behaviour is fundamental for energy balance purposes only in the transient phase. In fact, once this is finished, globally there will be no more effects due to the energy removed from the flow, because the phenomenon at the steady state is present in all the medium crossed by the heat flow, and therefore it will no longer give any contribution to the overall system budget. This is the physical explanation of why Cattaneo's equation gives way to the Fourier equation once the transient phase has elapsed.

Strictly connected to the duality of the system and to the energy conservation it is very instructive to recover the approach used by Baggioli et al. [22-25,20] to build up the Lagrangian describing systems (exhibiting *k*-gap) constituted by two mutually interacting sub-systems, that we identify in the DML with the *lattice particles* and the *liquid particles*. They introduce a two-fields potential representing displacements and velocities, $\phi_1$ and $\phi_2$, respectively. Neglecting the details of the mathematical formalism, what matters is that the equations of motion for the two scalar fields decouple, leading to two separate Cattaneo-like equations for $\phi_1$ and $\phi_2$, whose solutions are:

30.
$$\begin{cases} \phi_1 = \phi_0 \exp\left(-\dfrac{t}{2\langle \tau \rangle}\right)\cos(kx - \omega t) \\ \phi_2 = \phi_0 \exp\left(\dfrac{t}{2\langle \tau \rangle}\right)\cos(kx - \omega t) \end{cases}$$

Some consequences of this approach for the DML have been extensively discussed in [20]. Here-to-follow we want only to highlight those relevant for the topic dealt with in this paper. In particular, the interaction potential related to the Lagrangian is an oscillating function (see Figure 5 in [22] and Figure 2 in [24]), i.e. $\phi_1$ and $\phi_2$ reduce and grow over time $\langle \tau \rangle$, respectively (as electrical current and voltage do for an inductance); this implies that the two interacting sub-systems, represented by the two scalar fields, exchange among them energy and momentum, like wave-packets and *liquid particles* do. Therefore we may hypothesize that the interaction between the population of wave-packets and that of *liquid particles* is described by such couple of mutually interacting potentials. Besides, because the total scalar field is the product of $\phi_1$ and $\phi_2$, the total energy of the whole system does not vary with time, i.e. it is a constant of motion, as expected in the DML for systems constituted by the two populations of mutually interacting sub-systems (as in an electrical circuitry without sources or sinks). Finally, the motion is a typical dissipative hydrodynamic motion, and could represent a possible microscopic origin for the viscosity. Theories in which the system interacts with its environment have been used to explain important effects involving dissipation [22].

As pointed out by many authors [45-47], complex systems as liquids exhibit a distribution of relaxation times rather than one. This may be correlated also with the fact that in the DML [20-21] the propagation of energy (and momentum) can take place either as foreseen by the classical theory, that is by means of collisions between single molecules, and by means of interactions between the *lattice particles* and the *liquid particles*. In this second case, the propagation occurs through inelastic collisions between the wave packets and the molecule clusters, the ubiquitous presence of which in turn involves that of wave-packets, or *lattice particles*, by means of which the largest part of, if not quite all, the elastic energy and momentum are carried through and in between the icebergs. Besides, and this is also important to the validity of the model, the number of *lattice particles* has been shown to be comparable to that of the liquid molecules [20]. In [21] it has also been shown that the Eq.(13) for the specific heat $C_l^{wp}$ in DML as function of the collective DoF is in line with the experimental results, and its relevance with respect to the total specific heat, $C_V$, i.e. $C_V^{wp} \leq C_V$, has been extensively discussed. The classical treatment of heat propagation (see Appendix I) has always been applied to the molecular part alone, that is, the one described by the mutual interactions among the molecules of the liquid. The various authors have shown how the paradox of infinite propagation speed is resolved by introducing the delay term $\langle \tau \rangle$ in the diffusion equation, making it a propagation equation. In this work we have shown that even for the part of energy that propagates by means of the interactions of wave-packets with the pseudo-crystalline structures, heat propagation is described by a Cattaneo-like equation with "memory". While classically it is not possible to provide an explanation of the delay term and mainly of the undulatory behaviour in the transient phase, in the DML the introduction of the delay term has a clear physical explanation due to the tunnel effect, and the undulatory behaviour is made possible by the presence of the phonons as energy carriers. However, a formal generalization of the propagation equation, including either the part related to the wave-packets and the "classical" one related to molecular interactions, may be advanced, namely:

$$31. \quad \begin{cases} mol \quad J_q^{mol} + \left\langle \vartheta^{mol} \right\rangle \dfrac{\partial J_q^{mol}}{\partial t} = -K^{mol} \dfrac{\partial T}{\partial z} \\[4mm] wp \quad J_q^{wp} + \left\langle \vartheta^{wp} \right\rangle \dfrac{\partial J_q^{wp}}{\partial t} = -K^{wp} \dfrac{\partial T}{\partial z} \end{cases}$$

each one with own relaxation time and thermal conductivity. Because the internal energy of the system is distributed between the two sub-systems, one should consider an energy balance equation for the whole system. However, what really matters is that the thermal evolution of the system is characterized by two wave equations, each one with own medium parameters, relaxation time and, of course, wave propagation velocity, eq.(18). Indeed we have to do with a Dual System, the situation recalls that proposed by Landau for the supefluidity of HeII [27], i.e. that below the $\lambda-$ point liquid He is made either by ordinary and superfluid He, each one with own sound velocity. Equation (31$mol$) is associated with the macroscopic sound velocity, while eq.(31$wp$) is associated with the hyperfrequency sound velocity detected at mesoscopic scale for the first time by Ruocco and Sette [30].

Let's now return to the problem of the relaxation times. With reference to Eqs.(5) and (6) and to Figure 1a, the energy lost by the wave-packet is commuted into kinetic (i.e. translational) and potential (i.e. collective vibrations) energies of the iceberg, a relaxation time being definable for each process. The kinetic part of the particle' energy, involving external DoF, will be dissipated against the liquid friction with a own decay time $\tau_k$, along the decay process

32.    $\Delta E_k^p(t) = \Delta E_k^p(t=0) \cdot e^{-t/\tau_k}$ .

The part responsible for the excitation of the internal collective vibratory quantized DoF will be relaxed along with the relaxation time(s) $\tau_{\Psi_i}$ pertaining to the several excited distinct DoF, and following the decay processes [45]

33.    $\Delta \Psi_i^p(t) = \Delta \Psi_i^p(t=0) \cdot e^{-t/\tau_{\Psi_i}}$ .

This event is repeated many times as in a chain, giving origin to the thermal avalanche. To the purpose of the present paper, however, we still consider here a single relaxation time $\langle \tau \rangle$ (or $\langle \vartheta \rangle$), working for our interactions because what is relevant here is not just to distinguish between the several relaxation times one may consider, but rather establish that *there is* a delay physically associated to the *lattice particle $\leftrightarrow$ liquid particle* interaction at the origin of $\langle \vartheta \rangle$ in Eq.(15). This fact *per se* represents one of the novelties of this model, because the delay term in the transient phase in the DML represents a physical consequence of the model.

Eq.(15) solves the paradox of infinite propagation velocity of a thermal signal; in fact its solutions predict for heat transport a phenomenology of the diffusive type after an initial short-lived undulatory phase consisting in a wave propagation with velocity $\upsilon_C^{wp}$ given by eq.(18). By comparing Eqs. (18), (21), (28) and (29), we may easily recover the expressions for some well known quantities defined in irreversible thermodynamics (independently on whether the system is in propagative or diffusive (i.e. steady state) regime, [25]):

34. $\quad \upsilon_C^{wp} = \dfrac{\langle \Lambda_C \rangle_M}{\langle \vartheta \rangle_M} = \left( \dfrac{D_l^{wp}}{\langle \vartheta \rangle_M} \right)^{1/2} \Rightarrow D_l^{wp} = \dfrac{\langle \Lambda_C \rangle_M^2}{\langle \vartheta \rangle_M} = \upsilon_C \langle \Lambda_C \rangle_M$

We will show now under which conditions the picture given above provides the same results as that originally found by Cattaneo, in particular concerning the propagation velocity defined in Eq.(34). In the DML the wave-packets carrying thermal energy are seen as a gas of *lattice particles*, each of which moves with velocity $u^{wp}$ (we limit here to consider an idealized dispersionless elastic medium). In an isothermal system they move in random directions, so that we may consider the following relation holding for $u^{wp}$ [27,48-49]:

35. $\quad \left( u^{wp} \right)^2 = \left( u_x^{wp} \right)^2 + \left( u_y^{wp} \right)^2 + \left( u_z^{wp} \right)^2$

Therefore, the average RMS of the phonon velocity in any direction is:

36. $\quad \left\langle \left( \upsilon_C^{wp} \right)^2 \Big|_x \right\rangle = \left\langle \left( \upsilon_C^{wp} \right)^2 \Big|_y \right\rangle = \left\langle \left( \upsilon_C^{wp} \right)^2 \Big|_z \right\rangle = \left( \upsilon_C^{wp} \right)^2$

from which we finally get:

37. $\quad \upsilon_C^{wp} = \dfrac{u^{wp}}{\sqrt{3}}$

An energy disturbance in the phonon gas is carried through the liquid in the direction of propagation of the disturbance with the same velocity $\upsilon_C^{wp}$, which is related to the phonon velocity through Eq.(37), in line with the result get by Cattaneo [1].

Many authors have raised a number of objections about the use of a hyperbolic equations to describe the propagation of heat, especially i) for the physical meaning of the initial condition to be attributed to the temperature gradient, ii) because the total amount of heat, given by eq.(3) is no longer conserved, [50-51 and references therein] and iii) for their compatibility with II Principle. Here we do not want to enter into these debates, because what written in other works about the compatibility of hyperbolic equations with heat propagation is also valid in this case. However, we are keen to highlight the following aspects. One is connected to the first point. As widely explained before, in the DML there is a physical need for a propagative form of thermal energy in a liquid in the transient phase, rather than diffusive, due to the "tunnel" effect, by means of which energy is shifted from one place in the medium to another, Figure 1a; for each interaction, the length of the tunnel is $\langle \Lambda \rangle$ and the energy takes $\langle \tau \rangle$ to cross it. Let us then evaluate an expression one may easily deduce for the rate of temperature variation. To do this, we begin with considering the variation of heat flux $\Delta J_q^{wp}$ associated to wave-packets and due to a variation of the boundary conditions of the system. As outlined in section 2.2 and deeply discussed in [20], the current of wave-packets generates a pressure during its propagation, that in turn is equal to the amount of heat carried by phonons, $\Delta \Pi^{wp} = \Delta q_T^{wp}$. The overpressure wave $\Delta \Pi^{wp}$ multiplied by the velocity of advancement of the wave-front, $\upsilon_C^{wp}$, provides the power per unit cross-section, $\Delta J_q^{wp}$, dissipated by the wave-packets:

38. $\quad \Delta J_q^{wp} = \Delta q_T^{wp} \cdot \upsilon_C^{wp} = \dfrac{\partial q_T^{wp}}{\partial T} \Delta T \cdot \dfrac{\langle \Lambda_C \rangle}{\langle \vartheta \rangle}$

where $\Delta T$ is the local temperature increase due to $\Delta J_q^{wp}$. Dividing $\Delta J_q^{wp}$ by $\langle \Lambda_C \rangle$, one gets the thermal power per unit of volume dissipated by phonons along $\langle \Lambda_C \rangle$:

39. $\quad \Delta W_q^{wp} = \dfrac{\partial q_T^{wp}}{\partial T} \cdot \dfrac{\Delta T}{\langle \vartheta \rangle}$

Finally, dividing $\Delta W_q^{wp}$ by $C_V$, one gets the temperature variation vs time of the liquid contained in a volume of unitary cross-section and height $\langle \Lambda_C \rangle$:

40. $\quad \left\langle \dfrac{dT}{dt} \right\rangle = \dfrac{1}{\rho C_v} \dfrac{\partial q_T^{wp}}{\partial T} \cdot \dfrac{\Delta T}{\langle \vartheta \rangle} = \dfrac{\rho C_v^{wp}}{\rho C_V} \cdot \dfrac{\Delta T}{\langle \vartheta \rangle} = m \left[ \dfrac{m^*}{m^2} \dfrac{dm}{dT} T + 1 \right] \cdot \dfrac{\Delta T}{\langle \vartheta \rangle}$

where Eq.(13) for the phonon specific heat $C_l^{wp}$ has been used. Eq.(40) shows that the rate of temperature variation does not depend upon the external applied temperature gradient, but only on constitutive properties of the medium, in particular on the number of collective DoF excited.

A second remark concerns the energy conservation principle, that is of course not violated in the present approach. In fact, the missing part to the molecular energy content is that temporarily stored into non propagating DoF of the icebergs during its travel through the tunnel, that is not taken into account in dealing a system in the classical "single system" approach. This is another keypoint of the DML, which makes hyperbolic equations applicable to the energy (and mass) propagation in liquids due to the supposed "duality" of the system.

The final remark concerns the consistency of a propagation equation like Eq.(15) with the II Principle. In addition to what has already been written in support of this by various authors [7,14,16,19,46,52], the compatibility with the II Principle of thermodynamics has been dealt with in [53], in which an expression derived from a statistical approach is compared with one obtained from considerations of rational mechanics, where the non-isothermal fluid is considered as continuum. A modified stress tensor containing a term depending on heat flux can be obtained if one considers the system as a thermoelastic fluid, whose internal energy depends also on the heat flux. The insertion of this new independent variable in the energy equation leads to a modified entropy production and, in order for this model to be compatible with the II Principle, a Cattaneo type equation must be adopted to describe heat propagation.

## 4. Conclusions.

Contrary to its simplicity, the title of this work is by no means trivial, as often happens the most naïve questions are the most insidious and those requiring more attention. In this paper we have analyzed for the first time the problem of heat propagation in liquids modelled as Dual Systems, the two interacting subsystems being the population of phonons, or *lattice particles*, that interact with the population of *liquid molecules*, aggregates of liquid molecules, whose extension and density is function of temperature and pressure of the liquid. The two populations interact by means of $f^{th}$, that allows not only the propagation of energy but also of momentum. The interaction is characterized by a tunnel effect, by means of which the energy subtracted to the phonon pool is sequestered for a time lapse $\langle \tau \rangle$ in a non propagating form, the internal DoF of the liquid particle. Once $\langle \tau \rangle$ has elapsed, the liquid molecule has travelled through the liquid by $\langle \Lambda \rangle$ and in that point the energy emerges from the tunnel and returns to the phonon pool. This interaction has allowed to apply for the first time a hyperbolic equation to describe the heat (or mass) propagation in a liquid, thus solving the dilemma of infinite diffusion velocity which is typical of the Fourier law. The additional term distinguishing the Cattaneo type equation from the Fourier law is indeed justified from the physical point of view just by the anharmonic interaction *liquid particle* $\leftrightarrow$ *lattice particle*. Tunnelling in DML has various consequences. First, it represents the missing link to justify the use of a hyperbolic equation to describe the propagation of a thermal (and elastic) signal in a liquid. The hyperbolic equation is of course representative of the transient phase, when the effect of the propagation delay is also present at the macroscopic level. At stationary, the consequences of the tunnel effect are no longer appreciable at the macroscopic level, although still present at the microscopic level but equally distributed throughout the system. The hyperbolic equation reduces to the parabolic Fourier equation once the steady state is reached. Furthermore, the tunnel effect justifies the presence of shear waves on distances of the order of the dimensions of the molecular clusters. The above picture allows also to justify the additional initial condition needed to solve the hyperbolic equation, that results to be dependent only upon constitutive properties of the medium. It is therefore demonstrated another capability of DML.

DML offers of course many other perspectives of application to as many problems in the mesoscopic range of the physics of liquids. Future applications will certainly be addressed to compare our results with those of independent research groups. In particular, it is by far very interesting to get a way for calculating the values of the relaxation times involved in the heat propagation. Specific calculations are being finalized [28], although an Order-of-Magnitude has been already provided in this paper. In this frame it will be relevant to account for the works already published on this topic [7,50-51]. As well known, application of a temperature gradient to liquid solutions may reveal interesting phenomena [54], such as the selective diffusion of their components, the so-called thermo-diffusion. Provided that the system is gravitationally stabilized, the flux of thermal energy separates its components, until a dynamic equilibrium is reached, the Soret equilibrium. A similar, in some way opposite, phenomenon is observed if a gradient of concentration is applied; this generates a temperature difference inside the solution, the Dufour effect. Type (b) events of Figure 1 indeed show the capability of the reversible character of the elementary collisions, the particle that returns a more energetic phonon, undergoes a recoil as a result of the impulse conservation principle. This could explain at mesoscopic level the Dufour effect, i.e. the energy flux due to a mass concentration gradient occurring as a coupled effect of irreversible processes. The above listed phenomena are well-known and well described in non-equilibrium phenomenological thermodynamics. A possible future application for the DML will be that to provide a physical description of thermo-diffusion and mechano-thermal effects in terms of phonon $\leftrightarrow$ *liquid particle* inelastic collisions.

It is worth to mention the work by Zhao et al [55]. The authors use the expression for $C_V$ deduced in the Phonon Theory of Liquid Thermodynamics [56-57], a thermodynamic model of liquids in which these are assumed to be Dual Systems as in the DML. They develop a method to calculate the thermal conductivity of liquids, with the highest agreement with experimental data ever obtained, even if compared with previous models of liquids.

Experiments performed by Noirez and co-workers revealed an "unexpected" thermo-elastic effect in liquids confined at the mesoscopic level [58-60]. The mechano-thermal effect consists in the occurrence of a temperature gradient inside a liquid due to the momentum transferred by a moving plate to the liquid; this is confined in between one moving plate and one fixed. Although the thermal behaviour of the liquid in the several cases examined is not exactly the same, the dependence of the intensity of the mechano-thermal phenomenon, in particular of the temperature difference that originates, with respect to the distance between the two plates, remains almost the same. What matters here is that DML can provide a physical explanation for this "unexpected" phenomenon by means of the elementary wave-packet $\leftrightarrow$ *liquid particle* interaction, described in Figure 1b. The acceleration of the moving plate generates a momentum flux, directed towards the fixed plate. According to the DML, and also following Rayleigh's reasoning, the momentum flux, dimensionally a pressure, generates a temperature gradient. The *liquid particles*, pushed by the momentum flux, transfer their excess impulse to the phonons, which emerge from each single impact with an increased energy. They are then pushed and collected in proximity of the fixed plate, giving rise to a temperature gradient directed towards the fixed plate. The phenomenon is very rapid because it is determined by a succession of (almost) elastic collisions, which propagate very quickly (the relaxation time is of the order of $\langle \vartheta \rangle$). It is here important to highlight that the wave-packet $\leftrightarrow$ *liquid particle* collision also explains why a temperature gradient (i.e. a generalized flux-force crossed effect) is generated inside the liquid instead of a simple heating due to the energy stored as consequence of the mechanical stress $(G' >> G'')$, as one would have expected on the basis of a classical interpretation of the elastic modulus. In this framework it is interesting to look at eq.(40) although written in terms of pressure gradient instead of temperature. Eq.(40) tells us that, when a temperature (pressure) gradient is applied to a (dual) system, it generates a heat flux, the heat propagation being described by a hyperbolic equation (15). For such equation to be solved, one needs to specify an additional initial condition on temperature (pressure) variation vs time, quantitatively described by eq.(40). In the case of the experiments of Noirez we have to deal with a momentum gradient applied to a (dual) system; Figure 1b explains why it generates a temperature gradient as consequence of its propagation through the liquid. We propose here that an equation similar to eq.(40) could be setup to quantitatively justify the observed phenomena of the mechano-thermal effect. It is worth remembering that in eq.(40) only constitutive properties of the medium appear, in particular the number of collective DoF excited. A dedicated analysis will be done in a separate paper [28].

Trachenko and co-workers [61] have recently observed that the thermal diffusivity of liquids exhibit a lower bound that is fixed by fundamental physical constants for each system; even more intriguingly, the same value is shared by the kinematic viscosity of liquids. This in turn means that energy and momentum diffusion in liquids share the same universal lower bound and a strict correlation between the two transport mechanisms may be hypothesyzed. Interestingly, in the DML energy and momentum are both carried by interactions between the two subsystems constituting the liquids, *liquid particles* and *lattice particles*.

On the experimental side, performing experiments in non-stationary temperature gradients may allow to investigate how liquid parameters evolve from equilibrium to non equilibrium conditions. It could be possible for instance to investigate the evolution of correlations lengths, sound velocity, thermal conductivity, etc. A second type of experiments could be aimed at investigating the glassy and liquid-to-solid transitions. This last could be achieved starting from a stationary temperature gradient applied to a liquid sample and lowering both temperatures in order to lowering also the average temperature of the medium, until its solidification. Light scattering experiments performed during the non stationary phase should allow to investigate the dynamics of the system when the glassy and liquid-to-solid transitions are crossed. In particular, the external temperature gradient should orient the local domains along the same direction, thus allowing the increase of the correlation lengths, of sound velocity, of thermal conductivity, etc., along the preferential direction of the external temperature gradient. The experimental data collected should evidence a difference between the same parameters when measured along the direction of the temperature gradient with respect to those measured along a different direction should.

Also very interesting could be the investigation of whether a temperature gradient affects the viscous coupling between two adjacent liquids.

## 5.    Appendix. A few of history.

The history of Fourier's law modification is very long, and this appendix makes no claim to completeness. However, we wanted to bring to light what are perhaps the most salient points of the slow rise to prominence of the most widely used modification in mathematical physics, namely that of the Cattaneo equation, or in general, of the use of an equation with a linear term in the first time derivative of the heat flux. Still currently, the most referenced source, which collects most of the developments of the phenomenon of heat propagation (or diffusion), is represented by the two reviews by Joseph and Preziosi [50-51], which the reader is referred to if interested in detailed historical developments.

As well known, doubts about the validity of the Fourier equation for heat diffusion arose due to the infinite speed of heat propagation. The first to explicitly highlight the paradox was Cattaneo in his 1948 article [1]; the same issue was apparently addressed independently by Morse and Feshbach [62]. Surprisingly, the telegrapher' equation was however obtained for the first time by J.C. Maxwell [2] to describe the diffusion of heat in a material medium. Because he was interested in the diffusion phenomenon, he voluntarily eliminated the term containing the time derivative with the justification that "*The first term of this equation may be neglected, as the rate of conduction will* *rapidly establish itself*" (see [2], page 86 after eq. 143, and [3]). For this reason, no attention was paid to its importance in describing the diffusion process in its initial stages (as is known, the same equation can also be used to describe the diffusion of matter and not only of energy, giving rise to the Fick' equation). Maxwell was certainly right in his assumption, nevertheless, his approach contributed to keeping the issue hidden for quite a long time more.

Einstein, in his 1905 articles on Brownian motion [17-18], starting from a thermodynamic-statistical approach, establishes the relations that describe the diffusion coefficient in terms of known parameters of a solution; he is not concerned with the transient phase, but once again demonstrates how his model leads to the classical equation of steady-state diffusion.

In the early 1920's, Taylor [19] tackled the mathematical model of diffusion. He also deduces the classic Fick (or Fourier) equation since his goal is, once again, to describe the diffusion of particles at equilibrium, starting from a "discrete" model of how each individual particle advances.

To highlight the absurdity of the infinite velocity of propagation, Cattaneo [1] refers his analysis to the description made by Boltzmann [3] of a diffusive process. Cattaneo then introduces for the first time the concept of correlation between collisional events, and defines the relaxation time $\langle \tau_C \rangle$ in connection with the collision integral of the system, going so to demonstrate that the equation that correctly describes the diffusion is the one that will then take his name, and that he himself defines as an equation with "memory". Solutions of the hyperbolic equation predict for heat transport a phenomenology of the diffusive type, after an initial short-lived undulatory phase consisting in a wave propagation with velocity $\upsilon_C^{wp}$ (eq.37). However, no hint on the physical nature of these waves comes from Cattaneo approach. However, by digging into the depths of literature, one can discover how already in 1946 Peshkov [63] had hypothesized that in low temperature liquids "*a gas of thermal quanta capable of performing vibrations similar to those of sound should exist*".

The inconsistency of the Fourier' law with the Irreversible Thermodynamics was already highlighted by Onsager [8, pag. 419], because of the lacking of an acceleration term. A theory in which a thermal "inertia" is postulated is due to Kaliski [64], who generalized the Onsager relations and arrived at a telegraph equation after some simplifications. In 1951 Goldstein [14] retrieved Taylor's results [19] and Boltzmann's approach [3], and went, like Maxwell [2] and Cattaneo [1], to write a "telegrapher'" equation (without and with leakage) to describe the heat diffusion also in the transient phase.

It is also interesting to highlight how the various Authors mentioned above did not have a clear and full vision of the works of all their predecessors. Goldstein, for example, cites only Taylor in his work, and seems to ignore those of Maxwell, Boltzmann and Cattaneo. Cattaneo expressly cites Boltzmann's work, reusing it also in part, but ignores that of Maxwell. Boltzmann himself does not mention Maxwell's work in the bibliography of his book.

In 1958 Grad [65] unearths the additional time derivative term of eq.(15) plus a number of other corrections to the Fourier law. In 1963 M. Chester [15] resumed Maxwell's work, and he was the first to quote both Vernotte [4-5] and Cattaneo [6], but only the 1958 work of the latter, written in contrast to first Vernotte's paper, but not the original one [1]. In his interesting paper on the relaxation theory for thermal conduction in liquids, Nettleton [7] seems to know only the two papers by Vernotte, although he is well aware of the pioneering works by Frenkel [26], Brillouin [13] and Lucas [66]. Chester nevertheless uses Cattaneo's results to demonstrate that in dielectrics, or in solids without free electrons, there is a relationship between the speed of propagation of thermal energy deduced from the coefficients of the equation, and that of phonons, a relationship for which he provides a precise theoretical justification. Actually, Chester went much more deep in his reasoning about the heat propagation in condensed media. He indeed pointed out that the presence of second sound was due to the presence in the medium of phonons. Consequently, if second sound was foreseen, and then revealed in HeII, it could be detectable also in other media in which propagation of elastic, and thermal, energy was due to phonon propagation, such as non metallic solids, as well as liquids. The difference with HeII could only be the presence of a critical frequency for the appearance of second sound.

In 1982 Coleman, Fabrizio and Owen [16], in dealing with the consequences of using the Cattaneo equation to describe the propagation of the Second Sound in crystals, correctly quote Maxwell [2], Cattaneo [1] and Chester [3]. Joseph and Preziosi [50-51] collected in their articles a wide bibliography dedicated to the implications deriving from the various diffusion and propagative models. Very recently Ozorio Cassol and Dubljevic [29] compared parabolic and hyperbolic partial differential equations for heat diffusion. They cite both the 1958 work of Cattaneo [6] and that of Vernotte [4].

A special mention deserves a paper by Nettleton [67] in which the author deals with the density fluctuations in pure liquids by modelling the heat flux as made by two components, one carried by sound waves of very high frequency like elastic waves in a solid, $\psi_1$, and the other by molecules, $\psi_2$, so that $q = \psi_1 + \psi_2$. Both $\psi_1$ and $\psi_2$ will each obey a relaxation equation relating the time derivatives $\dot{\psi}_i$ $(i=1,2)$ to $\psi_i$ and $\nabla T$. At this point the author claims for the inertia of molecules to justify the fact that a sudden appearance of a temperature gradient will initially affect only $\dot{\psi}_i$ but not $\psi_i$. Then he builds a very general relaxation equation that he uses to obtain the density fluctuation and temperature dependence in the system. What matters of this approach is that i) Nettleton assumes the liquid thermal content as made of two components, and that ii) these two components interact among them in a similar way as the two two-fields potential representing displacements and velocities, $\phi_1$ and $\phi_2$, introduced in eq.(30), or alternatively, the two sub-systems described by eqs.(31).

**Funding:** This research received no external funding

**Data availability statement:** Not applicable

**Acknowledgements:** The author wants to express his gratitude to the Editor-in-Chief Prof. Enrico Bodo for the invitation to contribute to *Liquids* with an Invited Paper.

**Conflicts of interests:** The author declares no conflict of interests.

## 6. Figures

**Figure 1**

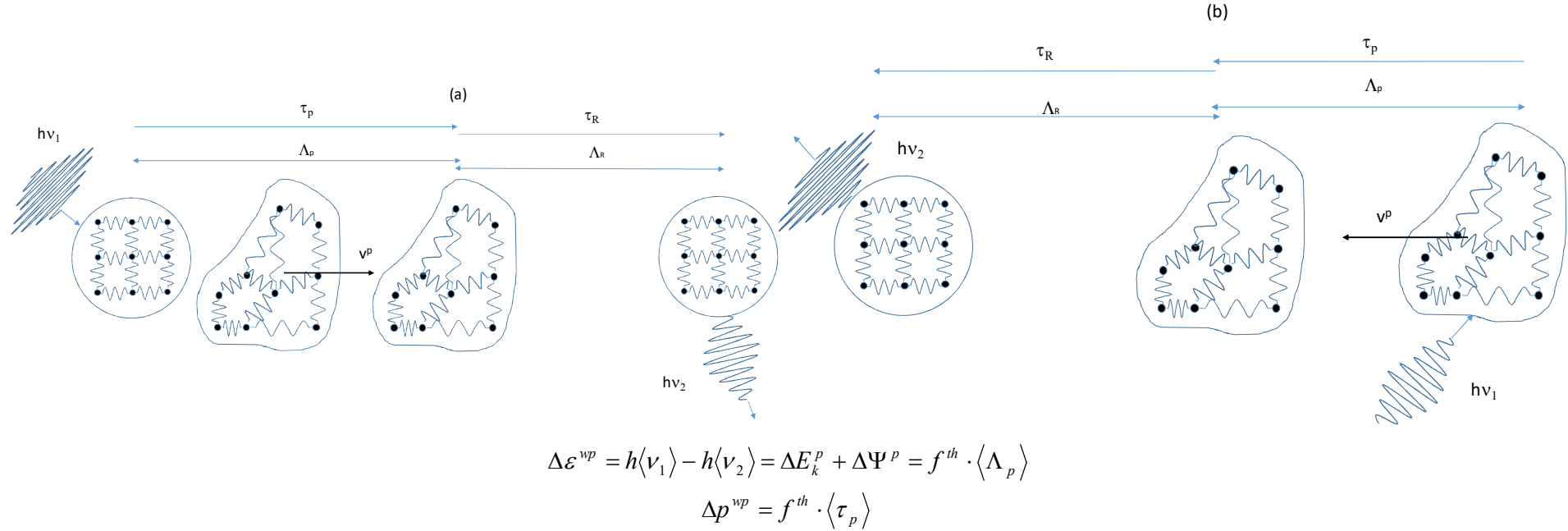

$$\Delta \varepsilon^{wp} = h\langle \nu_1 \rangle - h\langle \nu_2 \rangle = \Delta E_k^p + \Delta \Psi^p = f^{th} \cdot \langle \Lambda_p \rangle$$

$$\Delta p^{wp} = f^{th} \cdot \langle \tau_p \rangle$$

**Figure 1.** Schematic representation of inelastic collisions between wave-packets and liquid particles. The event represented in (a), in which an energetic wave-packet transfers energy and momentum to a liquid particle, is commuted upon time reversal into the one represented in (b), where a liquid particle transfers energy and momentum to a wave-packet. The particle changes velocity and the frequency of wave-packet is shifted by the amount $(\nu_2 - \nu_1)$. Due to its time symmetry, this mechanism has been assumed the equivalent of Onsager' reciprocity law at microscopic level [8-9,20].

In a pure isothermal liquid energy and momentum exchanged among the icebergs are statistically equivalent, and no net effects are produced. Events of type a) will alternate with events of type b), to maintain the balance of the two energy pools unaltered. Besides, the macroscopic equilibrium will ensure also the mesoscopic equilibrium; events (a) and (b) will be equally probable along any direction, to have a zero average over time and space. On the contrary, if a symmetry breaking is introduced, as for instance a temperature or a concentration gradient, one type of event will prevail over the other along a preferential direction.

(Re-drawn after Peluso, F., [21], MDPI Credits)

**Figure 2**

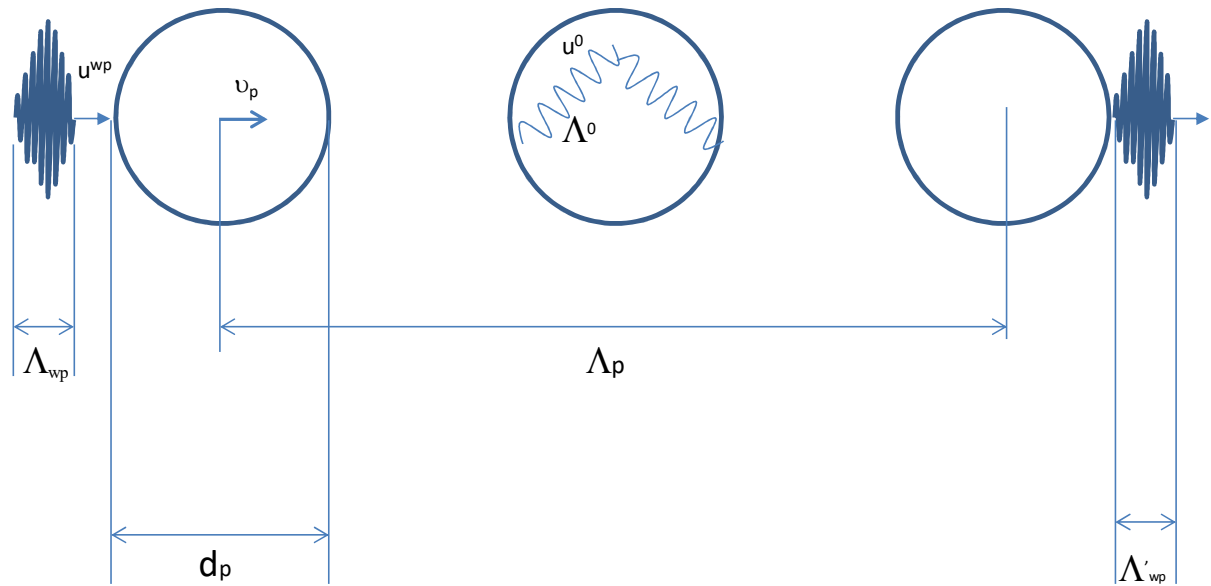

$$\left\langle \Delta^{ph} \right\rangle = \left\langle \Lambda_{wp} \right\rangle + \left\langle d_{p} \right\rangle + \left\langle \Lambda_{p} \right\rangle + \left\langle \Lambda'_{wp} \right\rangle$$

$$\left\langle \tau^{ph} \right\rangle = \frac{\left\langle \Lambda_{wp} \right\rangle}{u^{wp}} + \frac{d_{p}}{u^{0}} + \frac{\left\langle \Lambda_{p} \right\rangle}{u^{0}} + \frac{\left\langle \Lambda'_{wp} \right\rangle}{u^{wp}}$$

**Figure 2**. Close-up of the wave-packet – *liquid particle* interaction shown in **Figure 1**a. Only the first part of the interaction is represented, i.e. that during which the wave packet transfers momentum and energy to the liquid particle.

Modified after Peluso, F., see Reference [21]. MDPI Credits

# 7. References

1. Cattaneo, C., 1948, Sulla conduzione del calore, Atti Sem. Mat. Fis. Univ. Modena, **3**, 83-101

2. Maxwell, J.C., 1867, On the dynamical theory of gases, Ph. Trans. Royal Soc. London, **157**, 49-88.

3. Boltzmann, L., 1902, Leçons sur la théorie des gaz, Paris, Gauthiers-Villars, Ch. 2-11

4. Vernotte, P., 1958, Le paradoxe de la théorie continue de l'equation del la chaleur, Compt. Rend. Acad. Sci, **246**, 3154-3155.

5. Vernotte, P., 1958, La veritable équation de la chaleur, Compt. Rend. Acad. Sci, **248**, 2103-2105

6. Cattaneo, C., 1958, Sur une forme de l'equation de la chaleur éliminant le paradoxe d'une propagation instantanée, Compt. Rend. Acad. Sci, **247**, 431-433.

7. Nettleton.R.E., 1960, Relaxation theory of heat conduction in liquids, Phys. Fluids 3(2), 216-225

8. Onsager, L., 1931, "Reciprocal relations in irreversible processes I", Phys. Rev. 37, pp. 405-426

9. Onsager, L., 1931, Reciprocal relations in irreversible processes II", Phys. Rev, 38, pp. 2265-2279

10. Debye, P., 1912, Zur Theorie des specifische Wärmer, *Ann. der Physik*, **344** Nr.14, 798-839.

11. Debye, P., 1914, *Vorträge über die Kinetische Gastheorie*, pp 46-60, Leipzig, B.G. Teubner.

12. Brillouin, L., 1922, *Ann. de Phys.*, **17**, 88.

13. Brillouin, L., 1936, La chaleur spécifique des liquides et leur constitution, *J. Phys. Rad*, **Serie VII, Tome VII, N. 4**, 153-157

14. Goldstein, S., 1951, On diffusion by discontinuous movements, and on telegraph equation, Quart. Journ. Mech. and Applied Math, **4**, 129-156.

15. Chester, M., 1963, Second sound in solids, Phys. Rev, **131-5**, 2013-2015.

16. Coleman, B.D., Fabrizio, M., Owen, D.R., 1982, Il secondo suono nei cristalli: termodinamica ed equazioni costitutive, Rend. Sem. Mat. Un. Padova, **68**, 207-227.

17. Einstein, A., 1905, Investigation on the theory of Brownian movement, Ann. Phys., **17**, 549

18. Einstein, A., 1906, On the theory of Brownian movement, Ann. Phys., **19**, 371-381

19. Taylor, G.I., 1920, Diffusion by continuous movements, Proc. London Mat. Soc., **20**, 196-212.

20. Peluso, F. 2022, *Mesoscopic collective dynamics in liquids and the Dual Model*, ASME J. Heat Transfer, 144(11), 112502, https://doi.org/10.1115/1.4054988

21. Peluso, F., 2021 *Isochoric specific heat in the Dual Model of Liquids*, Liquids, 1(1), pp. 77-95; https://doi.org/10.3390/liquids1010007

22. Baggioli, M., et al., 2020, "Gapped Momentum States", Physics Reports, 865, 1-44.

23. Baggioli, M., Landry, M., Zaccone, A., 2022, "Deformations, relaxations and broken symmetries in liquids, solids and glasses: a unified topological theory", Phys. Rev. E, 105, 024602.

24. Baggioli, M., et al., 2020, Field Theory of Dissipative Systems with Gapped Momentum States, Phys. Rev. D, **102**, 025012, https://doi.org/10.1103/PhysRevD.102.025012.

25. Baggioli, M. Trachenko, K., 2019, Maxwell interpolation and close similarities between liquids and holographic models, Phs. Rev D, 99, 106002 doi: 10.1103/PhysRevD.99.106002

26. Frenkel, J., 1946, *Kinetic theory of liquids*, Oxford, Oxford University Press.

27. Landau, L.D., 1941, Theory of superfluidity of Helium II, Phys. Rev. 60, 356-358, doi: 10.1103/PhysRev.60.356

28. Peluso, F., In preparation

29. Ozorio Cassol, G., Dubljevic, S., 2019, Hyperbolicity of the Heat Equation, IFAC PapersOnLine, 52-7, 63-67, doi: 10.1016/j.ifacol.2019.07.011

30. Ruocco, G., Sette, F., Bergmann, U., Krisch, M., Masciovecchio, C., Mazzacurati, V., Signorelli, G., Verbeni, R., 1996, "Equivalence of the sound velocity in water and ice at mesoscopic lengths", Nature, 379, 521-523

31. Cunsolo, A., 2013, "Onset of a transverse dynamics in the THz spectrum of liquid water", Molecular Physics, 111(3), pp. 455-463, http://dx.doi.org/10.1080/00268976.2012.728258

32. Cunsolo, A., 2017, "The terahertz dynamics of simplest fluids probed by X-ray scattering", International Review in Physical Chemistry, 36-3, pp. 433-539.

33. Sette, F., Ruocco, G., Krisch, M., Bergmann, U., Masciovecchio, C., Mazzacurati, V., Signorelli, G., Verbeni, R., 1995, "Collective dynamics in water by high-energy resolution inelastic X-ray scattering", Phys. Rev. Lett., 75, pp. 850-854.

34. Sette, F., Ruocco, G., Krisch, M., Masciovecchio, C., Verbeni, R., 1996, "Collective dynamics in water by inelastic X-ray scattering", Physica Scripta, T66, pp. 48-56.

35. Sette, F., Ruocco, G., Krisch, M., Masciovecchio, C., Verbeni, R., Bergmann, U., 1996, "Transition from normal to fast sound in liquid water", Phys. Rev. Lett., 77, pp. 83-86

36. Ruocco, G., Sette, F., Krisch, M., Bergmann, U., Masciovecchio, C., Verbeni, R., 1996, "Line broadening in the collective dynamics of liquid and solid water", Phys. Rev. B, 54, pp. 14892-14895.

37. Sampoli, M., Ruocco, G., Sette F., 1997, "Mixing of longitudinal and transverse dynamics in liquid water", Phys. Rev. Lett., 79, pp. 1678-1681.

38. Sette, F., Krisch, M., Masciovecchio, C., Ruocco, G., Monaco, G., 1998, "Dynamics of glasses and glass-forming liquids studied by inelastic X-ray scattering", Science, 280, pp. 1550-1555.

39. Ruocco G., Sette, F., 1999, "The high-frequency dynamics of liquid water", J. Phys. Cond. Matt, 11, R259-R293.

40. Monaco G., Cunsolo A., Ruocco G., Sette F., 1999, "Viscoelastic behaviour of water in the THz frequency range: an inelastic X-ray study", Phys. Rev.E, 60-5, pp. 5505-5521.

41. Scopigno, T., Balucani, U., Ruocco, G., Sette, F., 2002, "Inelastic X-ray scattering and the high-frequency dynamics of disordered systems", Physica B, 318, pp. 341-349

42. Cunsolo A., Ruocco G., Sette F., Masciovecchio, C., Mermet, A., Monaco, G., Sampoli, M., Verbeni, R., 1999, "Experimental determination of the structural relaxation in liquid water", Phys. Rev. Lett., 82(4), pp. 775-778.

43. Cunsolo, A., 2017, "Inelastic X-Ray scattering as a probe of the transition between the hydrodynamic and the single-particle regimes in simple fluids", X-ray scattering, Ares, Alicia Esther Ed., chap. 1, Intech Open, http://dx.doi.org//10.5772/66126

44. Cunsolo, A., 2015, "The terahertz spectrum of density fluctuations of water: the viscoelastic regime", Adv Cond. Matt. Phys., 2015, pp. 137435-137459.

45. Herzfeld, K.F., Litovitz, T.A., 1959, *Absorption and dispersion of ultrasonic waves*, New York, Academic Press.

46. Nettleton, R.E., Compressional relaxation in liquids, *J. Acoust. Soc. Am.*, **31**, (1959) 557-567.

47. Nettleton, R.E, 1959, Thermodynamics of transport processes in liquids, *Trans. Soc. Rheol.*, **3**, 95-99.

48. Ward, J.C., Wilks, J., 1951, The velocity of second sound in liquid helium near the absolute zero, Phil. Mag. 42, 314-316 doi: 10.1080/14786445108561271.

49. Ward, J.C., Wilks, J., 1952, Second sound and the thermo-mechanical effect at very low temperatures, Phil. Mag. 43, 48-50, doi: 10.1080/14786440108520965

50. Joseph, D.D., Preziosi, L., 1989, Heat waves, Rev. Mod. Phys. **61**, 41

51. Joseph, D.D., Preziosi, L., 1989, Heat waves, Rev. Mod. Phys. **61**, 390

52. Gurtin, M.E., Pipkin, A.C., 1968, A general theory of heat conduction with finite wave speeds, Arch. Rational Mech. Anal. 31, 113-126.

53. Albanese, C., Mantile, A., Peluso, F., 2002, A new thermoelastic model for thermal radiation pressure, *Entropie*, **239-240**, 37-40

54. see for instance De Groot, S.R., Mazur, P., 1984, Non-equilibrum thermodynamics, New York, Dover pub. Inc..

55. Zhao, Z., Wingert, M.C., Chen, R., Garay, J.E., 2021, "Phonon gas model fro thermal conductivity of dense, strongly interacting liquids", J. Appl. Phys., 129, 235101; doi:10.1063/5.0040734

56. Bolmatov, D., Trachenko, K., 2011, "Liquid heat capacity in the approach from the solid state: Anharmonic theory", Phys. Rev. B, 84, pp. 054106(1-7).

57. Trachenko, K., Brazhkin, V.V., 2016, "Collective modes and thermodynamics of the liquid state", Rep. Prog. Phys. 79, pp. 016502-016538.

58. Kune, E., Zaccone, A., Noirez, L., 2021, "Unexpected thermoelastic effects in liquid glycerol by mechanical deformation", Phys. Flu., 33, 072007, doi: 10.1063/5.0051587

59. Kume, E., Noirez, L., 2021, "Identification of thermal response of mesoscopic liquids under mechanical excitations: from harmonic to nonharmonic thermal wave", J. Phys. Chem. B, 125, pp. 8652-8658, doi: 10.1021/acs.jpcb.1c04362

60. Noirez, L., Peluso, F., Private communication

61. Trachenko, K., et al., 2021, Universal lower bounds on energy and momentum diffusion in liquids, Phys. Rev. B 103, 014311, doi:10.1103/PhysRevB.103.014311.

62. Morse, P.M., Feshbach, H., 1953, *Methods of theoretical physics*, McGraw-Hill, New York, p. 865

63. Peshkov, V., 1946, in International Conference on Fundamental Particles and Low Temperatures, Cavendish Laboratory, Cambridge, July 22-27, 1946, Report p. 19-32, Taylor & Francis, London.

64. Kaliski, S., 1965, Wave equation of thermoelasticity, Bull. Acad. Pol. Sci, XIII (4), 253-260.

65. Grad, H., 1958, *Handbuch der Physik*, volume 12, p.271, edited by S. Flugge, Springer-Verlag, Berlin.

66. Lucas, R., 1938, "Sur l'agitation thermique des liquides, leur nouvelles propriétés thermomécahnique et leur conducibilité calorifique", J. Phys. 10, pp. 410-428.

67. Nettleton, R.E., 1960, Density Fluctuations and Heat Conduction in a Pure Liquid, Phys. Fluids, 4(1), 74-84, http://dx.doi.org/10.1063/1.1706190